\documentclass{aastex63}

\newcommand{\rev}[1]{#1}                    

\submitjournal{\apjs}
\accepted{December 20, 2020}

\graphicspath{{./}{figures/}}

\definecolor{tabblue}{rgb}{0.122, 0.467, 0.706}
\definecolor{taborange}{rgb}{1.0, 0.498, 0.055}
\definecolor{tabgreen}{rgb}{0.173, 0.627, 0.173}
\definecolor{tabred}{rgb}{0.839, 0.153, 0.157}
\newcommand{\textbfcol}[2]{\textbf{\textcolor{#1}{#2}}}

\usepackage[utf8]{inputenc}
\usepackage{graphicx}

\usepackage{savesym}
\savesymbol{tablenum}   
\usepackage{siunitx}

\usepackage{amsmath}    

\usepackage{url}
\usepackage{natbib}
\newcommand{\kepler}{\textsl{Kepler}}
\newcommand{\kebc}{\textsl{KEBC}}
\newcommand{\sebc}{\textsl{SEBC}}
\newcommand{\galaxia}{\texttt{Galaxia}}
\newcommand{\tess}{\textsl{TESS}}
\newcommand{\lsst}{\textsl{LSST}}
\newcommand{\vro}{\textsl{VRO}}
\newcommand{\python}{\texttt{Python}}
\newcommand{\ktwofov}{\texttt{K2fov}}
\newcommand{\gaia}{\textsl{Gaia}}
\newcommand{\plato}{\textsl{PLATO}}
\newcommand{\kspc}{\textsl{KSPC}}

\newcommand{\ssth}{\textsuperscript{th}}

\def\equationautorefname~#1\null{Eq.~(#1)\null}

\newcommand{\refeqn}[1]{Eq.~(#1)}

\DeclareSIUnit\msun{M_\sun}
\DeclareSIUnit\rsun{R_\sun}
\newcommand{\eccn}[0]{\ensuremath{e}}
\newcommand{\emax}[0]{\ensuremath{e_{\mathrm{max}}}}
\newcommand{\emaxMDS}[0]{\ensuremath{{\emax}_{,\mathrm{MDS}}}}
\newcommand{\logp}[1]{\ensuremath{\log{\!P_{#1}}}}
\DeclareSIUnit\sqdeg{deg^2}

\newcommand{\lage}[1]{\ensuremath{\log{\!A_{#1}}}}

\newcommand{\nucrit}[0]{\ensuremath{\nu_{\mathrm{crit}}}}

\newcommand{\sgn}{\operatorname{sgn}}

\begin{document}

\title{Building and Calibrating the Binary Star Population Using \kepler{} Data}


\author[0000-0003-1551-3717]{Mark A. Wells}
\affiliation{
    Department of Astrophysics \& Planetary Science,
    Villanova University,
    Villanova, PA 19085, USA
}
\affiliation{
    Department of Astronomy \& Astrophysics,
    Eberly College of Science,
    The Pennsylvania State University,
    University Park, PA 16802, USA
    }
\author[0000-0002-1913-0281]{Andrej Pr{\v s}a}
\affiliation{
    Department of Astrophysics \& Planetary Science,
    Villanova University,
    Villanova, PA 19085, USA
}

\begin{abstract}
Modeling binary star populations is critical to linking the theories of star formation and stellar evolution with observations.
In order to test these theories, we need accurate models of observable binary populations.
The \kepler{} Eclipsing Binary Catalog (\kebc{}), with its \rev{estimated} $>$90\% completeness, provides an observational anchor on binary population models.
In this work we present the results of a new forward-model of the binary star population in the \kepler\ field.
The forward-model takes a single star population from a model of the galaxy and pairs the stars into binaries by applying the constraints on the population from the results of observational binary population surveys such as \citet{Raghavan.etal.2010} and \citet{Duchene.Kraus.2013}.
A synthetic binary population is constructed from the initial distributions of orbital parameters.
We identify the eclipsing binary sample from the generated binary star population and compare this with the observed sample of eclipsing binaries contained in the \kebc{}.
Finally, we update \rev{the distributions} of the synthetic population and repeat the process until the synthetic eclipsing binary sample agrees with the \kebc{}.
The end result of this process is a model of the underlying binary star population that has been fit to observations.
We find that for fixed flat mass ratio and eccentricity input distributions, \rev{the binary} period distribution is logarithmically flat above $\sim$\SI{3.2}{d}.
With additional constraints on distributions from observations, we can further adjust the synthetic binary population by relaxing other input constraints, such as mass ratio and eccentricity.
\end{abstract} 

\section{Introduction}
Binary star populations are the products of star formation and stellar evolution within a stellar environment.
The distribution of their intrinsic parameters---multiplicity, mass ratio, period, and eccentricity---provide us with clues about the inner workings of these processes.
A viable model of a stellar environment's binary population yields, among others, insight into its star formation history (see \citet{Moe.DiStefano.2017} and references therein for an overview).
The creation and validation of such models would ideally utilize results from volume-limited surveys.
Unfortunately, most surveys (including \kepler{}) are magnitude-limited, and suffer from Malmquist bias where more massive and luminous systems are observed at a disproportionate rate to their occurrence.

Survey missions that observe large regions of the sky with repeat observations lend themselves readily to the study of binary stars through the detection of eclipses.
In particular, the \kepler{} mission \citep{Borucki.etal.2010} provided us with the most complete census of eclipsing binaries in its 105-deg$^2$ field of view \citep{Kirk.etal.2016}.
Ongoing missions such as the Transiting Exoplanet Survey Satellite \citep[\tess{};][]{Ricker.etal.2015} and upcoming surveys such as the Legacy Survey of Space and Time (\lsst) at the Vera Rubin Observatory \citep[\vro;][]{lsstSRD, Ivezi.etal.2008, Abell.etal.2009}, and the Planetary Transits and Oscillations of Stars Mission \citep[\plato{};][]{Magrin.etal.2018} will greatly expand upon the observations made by \kepler{}.
The \vro{}, capable of performing all-sky observations down to $r \sim 24.5$, will include the faintest and lowest-mass binary populations in the Galaxy that have thus far been largely unobserved.
With these surveys on the horizon, having a robust framework in place that allows for rapid utilization of the incoming data cannot be overstated.

Previous work has been done to include binarity in galaxy models.
\citet{Arenou.2011} used the Besan\c{c}on Galaxy Model \citep[BGM,][]{Robin.etal.2003} to simulate \gaia\ data.
\citet{Czekaj.etal.2014} did work to update the BGM, using the strategy laid out by \citet{Arenou.2011} to include binarity and compared their results against the Tycho-2 catalog \citep{Hog.etal.2000}.
While the BGM scheme developed in \citet{Czekaj.etal.2014} has the benefit of conserving the local stellar mass density, it uses fixed input distributions for mass ratio, period, and eccentricity.
Unfortunately, the fact that the underlying distributions can not be modified, as well as the work's unmaintained status, means that it is not well suited to our needs.

In this work we present the foundations of a binary population synthesis framework and demonstrate its application on the \kepler{} binary population.
We synthesize a binary population, generate a sample of eclipsing binaries, and compare these eclipsing binaries to the \kepler{} Eclipsing Binary Catalog \citep[\kebc{};][]{Prsa.etal.2011, Kirk.etal.2016}.
Using the \kebc{} as ``ground truth'', we update \rev{the binary population model} and re-synthesize.
This process is iterated until the generated eclipsing binary population has converged with the observed population.

\section{Methodology}
During the process of creating our synthetic binary star population, we build upon a number of works.
To model the galaxy, we use \galaxia{}\footnote{available at \url{http://galaxia.sourceforge.net/}} \citep{Sharma.etal.2011}, a code that produces a synthetic single star survey of the Milky Way.
\galaxia{} implements the Besan\c{c}on Milky Way disk model \citep{Robin.etal.2003}.
In addition, we borrow the shape of our multiplicity relationship from \citet{Arenou.2011} and modify it to be in better agreement with observations of multiplicity from \citet{Raghavan.etal.2010} and \citet{Duchene.Kraus.2013}.

A binary system consists of a more massive primary star and a less massive secondary star.
Their orbits are fully described by the following parameters: the mass of the primary, $M_1$; the mass ratio, $q=M_2/M_1$, where $M_2$ is the mass of the secondary; \rev{the period, $P$;} the eccentricity, $e$; the inclination, $i$; and the argument of periastron, $\omega$.
Each star from \galaxia{} is assigned a probability of either serving as a primary of a binary system or of being a single star.
This probability is determined by our adopted multiplicity relationship.
The multiplicity fraction is the ratio of multiple systems (binaries and higher order multiples) over all systems (singles and multiples) and is a function of primary mass.
Once the primary stars are selected, the orbital parameters and secondary stars are drawn according to our model.
The model currently takes $q$ and $e$ as static input distributions while \logp\ uses a discrete distribution.
In this work we limit ourselves to detached binaries as contact binaries would require additional evolutionary models.
We draw orbital parameters with the additional requirements that the systems are well detached.
Once we have fully specified binaries, we compute the observing geometries of the synthetic binaries and determine which systems will present an eclipse.

There are several observational biases that need to be accounted for before we can compare against actual observation.
In order to simulate the target selection process, we draw synthetic targets that have similar magnitudes and color to the actual \kepler\ target list.
By using only magnitudes and color, we are mimicking the actual process that was used to determine the initial target list \citep{Batalha.etal.2010}.
Our method of target selection uses no defined relationships but simply selects a sample that most closely resembles the empirical  distribution of the \kepler\ targets.
We create a model of the \rev{\kepler\ detection efficiency as a function of period} which is then used to determine the fraction of observed eclipsing binaries.

We compare the resulting period distribution of the synthetic eclipsing binaries directly to the observed \kepler\ Eclipsing Binary Catalog.
Based on the ratio of the densities of the synthetic and observed distributions, we calculate a set of corrections to apply to the initial \logp\ distribution.
These corrections are used in turn to synthesize a new, slightly different, synthetic population of binaries.
The process of adjustment and synthesis is repeated until the model has settled about a solution.
The number of iterations is chosen such that the simulated eclipsing binary catalog fluctuates about the \kebc{}.
The final 50~runs are used to compute the mean and standard deviation for each bin of the output distributions.

\subsection{Generating the Stellar Population} \label{sec:genstelpop}
\begin{figure}
    \includegraphics{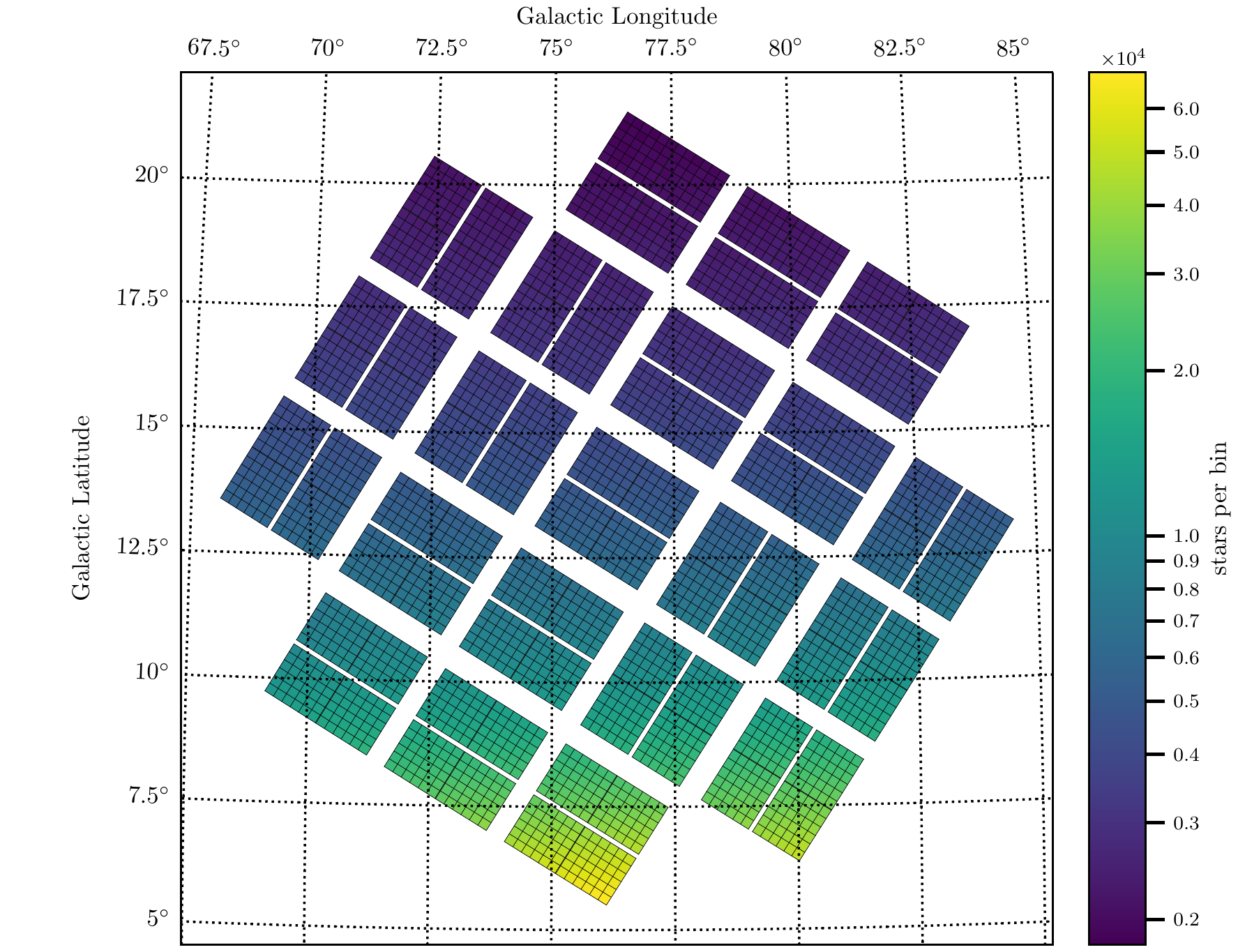}
    \caption{
        Color map of the synthetic stellar population (down to 25\ssth\ magnitude), created with \galaxia, of the \SI{115}{\sqdeg} \kepler\ field of view plotted in galactic coordinates.
        The number density of stars increases as the line of sight moves toward the galactic disk.
    }
    \label{fig:kepfield-glonlat}
\end{figure}

\begin{deluxetable*}{lllRR}
    \tablecaption{\label{tab:galsettings}
        \galaxia\ settings used to generate the stellar population.
        The table contains the name of each setting, a brief description, the default value from \galaxia\ and the value that we used.
    }
    \tablehead{\colhead{parameter} & \multicolumn2c{Description} & \colhead{Default} & \colhead{Value}}
    \startdata
    \texttt{photoSys} & \multicolumn{2}{l}{photometry system to use} & UBV\tablenotemark{a} & SDSS\tablenotemark{b}\\
    \texttt{magColorNames} & \multicolumn{2}{l}{arguments for \texttt{appMagLimits}, \texttt{colorLimits}} & $V$, $B-V$ & $r$, $g-r$\\
    \texttt{appMagLimits[0]} & \multicolumn{2}{l}{apparent magnitude lower limit} & -100 & 100\\
    \texttt{appMagLimits[1]} & \multicolumn{2}{l}{apparent magnitude upper limit} & 30 & 25\\
    \texttt{absMagLimits[0]} & \multicolumn{2}{l}{absolute magnitude lower limit} & -100 & -100\\
    \texttt{absMagLimits[1]} & \multicolumn{2}{l}{absolute magnitude upper limit} & 100 & 100\\
    \texttt{colorLimits[0]} & \multicolumn{2}{l}{lower limit on color} & -100 & -100\\
    \texttt{colorLimits[1]} & \multicolumn{2}{l}{lower limit on color} & 100 & 100\\
    \texttt{geometryOption} & \multicolumn{2}{l}{0: all sky,  1: circular patch} & 1 & 1\\
    \texttt{longitude}\tablenotemark{c} & \multicolumn{2}{l}{galactic longitude of circular patch center [deg]} & 0 & *\tablenotemark{d}\\
    \texttt{latitude}\tablenotemark{c} & \multicolumn{2}{l}{galactic latitude of circular patch center [deg]} & 90 & *\tablenotemark{d}\\
    \texttt{surveyArea}\tablenotemark{c} & \multicolumn{2}{l}{survey area [sq deg]} & 100 & *\tablenotemark{d}\\
    \texttt{fSample} & \multicolumn{2}{l}{fraction of stars to generate} & 1.0 & 1.0\\
    \texttt{popID} & \multicolumn{2}{l}{ID of population to generate ($-1$ is all populations 0-9)} & -1 & -1\\
                   & 0: thin disk $<\SI{0.15}{Gyr}$ & 1: thin disk \SI{0.15}{Gyr} to \SI{1}{Gyr} & & \\
                   & 2: thin disk \SI{1}{Gyr} to \SI{2}{Gyr} & 3: thin disk \SI{2}{Gyr} to \SI{3}{Gyr} & & \\
                   & 4: thin disk \SI{3}{Gyr} to \SI{5}{Gyr} & 5: thin disk \SI{5}{Gyr} to \SI{7}{Gyr} & & \\
                   & 6: thin disk \SI{7}{Gyr} to \SI{10}{Gyr} & 7: thick disk  & & \\
                   & 8: stellar halo & 9: bulge & & \\
                   &\multicolumn{2}{l}{10: \citet{Bullock.Johnston.2005} stellar halos} & & \\
    \texttt{warpFlareOn} & \multicolumn{2}{l}{warp and flare the thin disk (0: no, 1: yes)} & 1 & 1\\
    \texttt{rmax} & \multicolumn{2}{l}{maximum radial distance [kpc]} & 1000 & 1000\\
    \enddata
    \tablenotetext{a}{\citet{Johnson1953}}
    \tablenotetext{b}{\citet{Fukugita1996}}
    \tablenotetext{c}{Used only if $\texttt{geometryOption}=1$}
    \tablenotetext{d}{Varies for each \kepler{} module.}
\end{deluxetable*}
\galaxia{} takes a variety of parameters that fine-tune the generated population, as shown in \autoref{tab:galsettings}.
Of all these parameters, we will only focus on the magnitude limits and sky position as these are unique to our process.
We generate stars down to a magnitude of $r_\mathrm{SDSS} = 25$.
While this is roughly 5 orders of magnitude fainter than the magnitude limit of \kepler{}, it is required in order to generate the low-mass, and subsequently faint, stellar components that will become secondaries.
We create circular regions that are centered on, and circumscribe, each of the 22 \kepler{} modules.
These synthetic stellar populations are then trimmed using the \python{} package \ktwofov{}\footnote{Available at \url{https://github.com/KeplerGO/K2fov}.} \citep{k2fov} to identify which objects actually fall on the active silicon of the detector.
The 22 stellar populations, now trimmed, are combined and constitute our population of single stars (cf. Figure~\ref{fig:kepfield-glonlat}).
%

\subsection{Drawing Primaries} \label{sec:drawprims}

\begin{figure}[htb!]
    \includegraphics[]{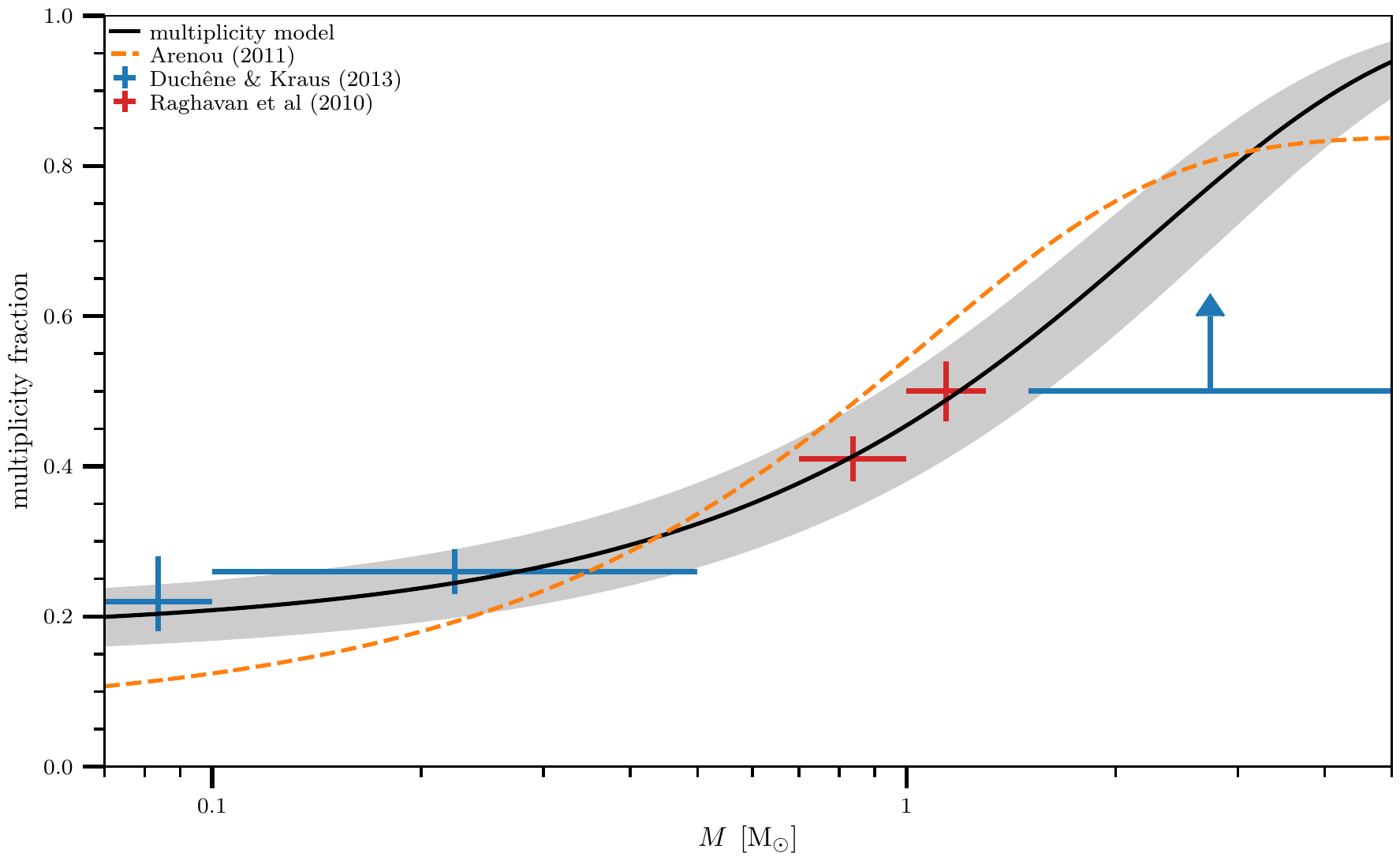}
    \caption{
        We adopt the form of our multiplicity model (\textbf{black line}) from \citet{Arenou.2011} (\textbf{gray dashed line}) and modify it to better fit the results of \citet{Duchene.Kraus.2013} and \citet{Raghavan.etal.2010}.
    }
    \label{fig:mult-frac}
\end{figure}

The first step in synthesizing the binaries is to select a primary star for each system.
Primaries are selected from the single star population in accordance with their respective multiplicity fraction.
The multiplicity fraction is defined as the ratio:
\begin{equation}
    f_m \equiv \frac{B+T+\ldots{}}{S+B+T+\ldots}
\end{equation}
where $S$, $B$, $T$, \dots are the numbers of single, binary, triple and higher-order multiple systems, respectively.
For a star of mass $M$ in the stellar population, we say it has a probability of $f_m$ to be drawn as a primary star.
This relationship depends on mass and we adopt the following analytical relationship from \citet{Arenou.2011}:
\rev{%
\begin{equation}
    f_m(M) = c_1 \tanh(c_2 M + c_3)
\end{equation}
where $c_1$, $c_2$, and $c_3$ are free parameters and $M$ is the stellar mass.
We modify the original coefficients given by \citet{Arenou.2011} to better fit data reported by \citet{Duchene.Kraus.2013} and \citet{Raghavan.etal.2010} (see the top panel of Figure~\ref{fig:mult-frac}).
The coefficients given by \citet{Arenou.2011} are $c_1=0.8388$, $c_2=0.688$, and $c_3=0.079$, while our modified values are $c_1=1$, $c_2=0.31\pm0.05$, and $c_3=0.18\pm0.04$.
We have fixed $c_1$ to unity as only $6_{-3}^{+6}\%$ of O-type stars are singles \citep{Moe.DiStefano.2017}.
}%
Due to the inherent simplicity of the analytical model assumed here, our derived relationship potentially overestimates the multiplicity rates above \SI{3}{\msun}.
We expect this to be a marginal issue for the \kepler\ field as approximately only 400 single stars, out of 28.7~million, have masses above \SI{3}{\msun},
In addition, even if all of these systems, single stars and binaries with primaries above \SI{3}{\msun}, made it into the target list, this would only make up approximately 0.2\% of the observed systems.
For higher mass populations, a better analytical model will need to be considered.
After the primary stars have been drawn, we turn our attention to the orbital parameters.

\subsection{Orbital Parameters and Secondaries} \label{sec:orbparams}

\begin{figure}[htb!]
    \includegraphics[]{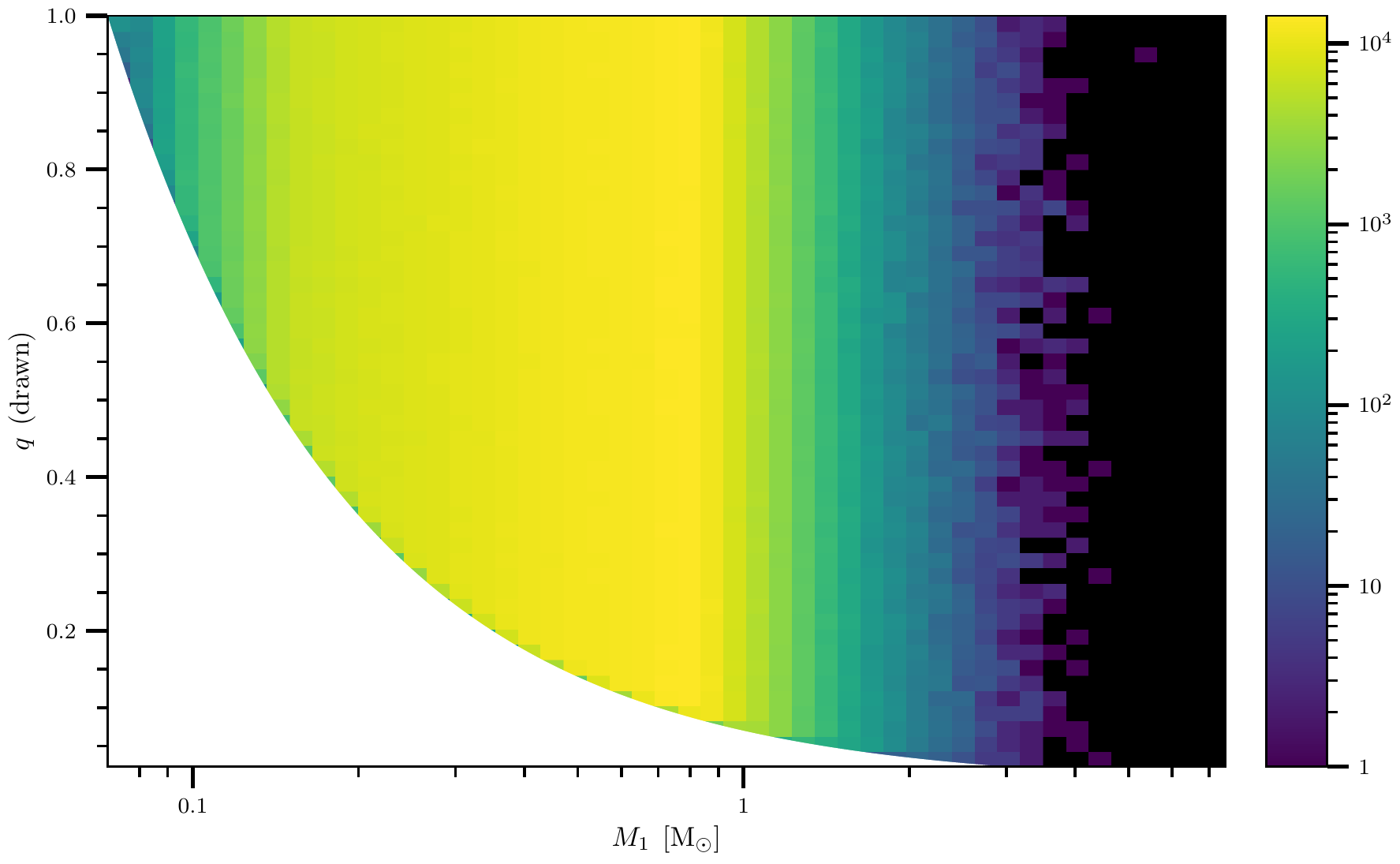}
    \caption{
        Color map of the number of systems as a function of mass ratio ($q$) and primary mass $M_1$.
        The lower left portion of the map is forbidden by our criterion on $q$ stated in \autoref{eqn:q0}.
        The change in intensity along $M_1$ is reflective of two competing factors.
        The decrease that occurs at $M_1 \gtrsim \SI{1}{\msun}$ is due to the declining abundance of high mass stars, while the decrease that occurs at $M_1 \lesssim \SI{0.2}{\msun}$ is due to the Malmquist bias of our generated sample.
    }
    \label{fig:mass-mrat}
\end{figure}

Binary star orbits are characterized by the mass ratio ($q = M_2/M_1$), orbital period ($\log P$ with $P$ in days), orbital eccentricity ($e$), semi-major axis ($a$), inclination ($i$), and the argument of periastron ($\omega$).
It is necessary to place strict limits on mass ratio and period to ensure that only viable systems are generated.
Mass ratio, along with the mass of the primary, determines the mass of the secondary.
The minimum mass for a star to sustain nuclear fusion is taken to be \SI{0.07}{\msun}, chosen to agree with \galaxia.
This sets a limit on the mass ratio, as a function of primary mass, given by:
\begin{equation}
    \label{eqn:q0}
    q_{\min}(M_1) = \frac{\SI{0.07}{\msun}}{M_1}.
\end{equation}
The relationship between mass ratio and primary mass is illustrated in \autoref{fig:mass-mrat}.
The truncation of the distribution is clearly visible and is more pronounced for lower mass stars.
After the mass ratio has been drawn, we can draw secondary stars.

In addition to the mass requirement set by mass ratio, we also require the stars to be coeval.
To determine the most appropriate secondary star from the synthetic star population to serve as a secondary we use the following merit function for each binary
\begin{equation}
    \label{eqn:secmetric}
    S(M,\lage{}) = \sqrt{\left(\frac{M - qM_1}{M_{\max} - M_{\min}}\right)^2 + \left(\frac{\lage{}-\lage{1}}{\lage{\max} - \lage{\min}}\right)^2}
\end{equation}
where $S$ is our secondary star merit function, $M_{\min}$, $M_{\max}$, \lage{\min}, and \lage{\max} are the minimum and maximum masses (in solar units) and ages (in Gyr) of the synthetic star sample, respectively.
We divide by the difference of the extrema to place the parameters on a normalized scale.
A secondary is selected if the value of $S$ does not exceed 0.01 and if $M<M_1$.
We include the condition $M<M_1$ to ensure that our primaries are the more massive members of the system.
The value of 0.01 corresponds to a combined normalized parameter deviation of 1\% from the desired mass of $qM_1$ and \lage{1}.
The minimum and maximum values of mass and age used are $M_{\min}=\SI{0.07}{\msun}$, $M_{\max}=\SI{7.49}{\msun}$, $\lage{\min}=4.94$, and $\lage{\max}=10.1$.
Therefore the selected secondary is guaranteed to have $\left|M_2 - qM_1\right| \le \SI{0.07}{\msun}$ and $\left|\lage{2} - \lage{1}\right| \le 0.05$.

The drawn secondary adopts the non-stellar properties of the primary such as location in the sky, distance, and interstellar extinction.
This process does not preserve stellar mass content as secondaries are drawn with replacement.
The resulting synthesized populated of single stars and binaries is inflated relative to the model stellar population produced with \galaxia.

Once secondaries have been drawn for each system, we can draw period (\logp{}) and eccentricity ($e$).
To model the \logp{} distribution, we construct a binned model, with $n$ bins of equal width and an additional overflow bin.
The $n$ bins span the observable eclipsing binary \logp{} range of -0.64 (5.5~hours) to 2.86 (730~days).
Of course binaries exist outside of this range and, to account for this, we have included an additional bin which ranges up to 8.0 ($\approx \SI{270}{kyr}$).
While the $n$ bins that fall within the observable range are fit explicitly, the overflow bin is fit by requiring the integral of the model density to be unity.
Hence, the overflow bin scales the rest of the model and is an estimate of how many binaries are above the maximum observable period.

Before we can draw from the period model, we must first determine the minimum allowed period, $\logp{\min}$.
To ensure that we only generate detached binaries, we filter the separation of the system at periastron with:
\begin{equation}
    \label{eqn:r1r2}
    a(1 - e) > s\left(R_1 + R_2\right),
\end{equation}
where $a$ is the semi-major axis, $s$ is the separation factor, $R_1$ is the radius of the primary, and $R_2$ is the radius of the secondary.
The separation factor in \autoref{eqn:r1r2} must be set to a value above 1 to ensure that adequate separation between the two stars is maintained.
A star that is critically rotating, such that it can not spin faster without ejecting mass, will have its equatorial radius equal to $\frac{3}{2}$ of its polar radius (see \citet{Prsa2018} for a complete derivation).
We have chosen to set $s=\frac{3}{2}$ to correspond to the situation in which we have both stars spinning at their critical breakup rotation.
The minimum separation of the system:
\begin{equation*}
    a_{\min} = \frac{3(R_1+R_2)}{2(1-e)}
\end{equation*}
is used, along with Kepler's 3rd law, to obtain:
\begin{equation*}
    P_{\min} = 3\pi\left(\frac{3}{2G(M_1+M_2)}\right)^{\frac{1}{2}} \left(\frac{R_1+R_2}{\vphantom{|}1-e}\right)^{\frac{3}{2}}.
\end{equation*}
The period is drawn from the distribution with an additional requirement that $\logp{} \ge \logp{\min}$.

Eccentricity is uniformly drawn between 0 and \emax{} and occurs only after period has been drawn.
We use two criteria to determine \emax: the first given by \autoref{eqn:r1r2} and the second from \refeqn{3} of \citet{Moe.DiStefano.2017}.
The criterion from \citet{Moe.DiStefano.2017},
\begin{equation*}
    \emaxMDS = \begin{cases}
        0 & \text{for } P \leq \SI{2}{d} \\
        1 - \left(\frac{P}{\SI{2}{d}}\right)^{-2/3} & \text{for } P > \SI{2}{d}
    \end{cases}
\end{equation*}
assumes everything below \SI{2}{d} is circularized and guarantees that the binary components do not fill their Roche lobes by more than 70\%.
Combining this with \autoref{eqn:r1r2}, we have:
\begin{equation}
    \emax = \min\left(1 - \frac{1.5(R_1+R_2)}{a},\ {\emax}_{,\mathrm{MDS}}\right)
    \label{eqn:emax}
\end{equation}
where we use whichever criterion provides the more conservative value for \emax.

The final properties to be drawn are inclination and argument of periastron.
Inclination is sampled uniformly in terms of $\cos{i}$ for $i$ between \SI{0}{\degree} and \SI{180}{\degree}.
The argument of periastron, $\omega$, is drawn uniformly between \SI{0}{\degree} and \SI{360}{\degree}.

\subsection{Eclipsing Binaries}

Once we have a sample of synthesized binary systems, we need to determine which of those will eclipse.
We compute the \rev{projected} separation of the system and use
\begin{equation*}
    R_1+R_2>r(\nu)\cos{i}
\end{equation*}
as the criterion for an eclipse where $r(\nu)$ is the instantaneous separation of the stellar centers, as a function of true anomaly, given by
\begin{equation*}
    r(\nu) = \frac{a(1-e^2)}{1+e\cos(\nu)}.
\end{equation*}
For circular orbits, there are two critical points, \nucrit{}, that correspond to the minimum projected separation of the system.
These points occur at $\nucrit{} = \frac{\pi}{2} - \omega$ and $\nucrit{} = \frac{3\pi}{2} - \omega$.
For eccentric orbits the point of projected closest approach cannot be solved analytically.
The criterion we use to determine if a system is a synthetic eclipsing binary is
\rev{%
\begin{equation}
    R_1+R_2 > \frac{a (1-e^2)}{1+e \cos(\nucrit{})}\sqrt{\cos^2(\nucrit{} + \omega) + \sin^2(\nucrit{} + \omega)\cos^2i}
    \label{eqn:eclip-geom}
\end{equation}
where the right side of \autoref{eqn:eclip-geom} is the projected separation of the system in terms of the orbital parameters.
}%
We evaluate \autoref{eqn:eclip-geom} at both values of \nucrit\ to determine if an eclipsing event occurs.

Currently, we do not differentiate between binaries which have only a single eclipsing component versus both components.
Our eclipse criterion assumes spherical stars, which is appropriate given that all systems have been generated to ensure that they are detached.
In addition, we do not consider the depth, nor the profile, of the eclipsing signal which would require a \kepler\ light-curve noise model and a limb darkening model, respectively.
The systems that satisfy the geometry constraint, \autoref{eqn:eclip-geom}, are considered to be eclipsing.


\section{Observational Effects}
\subsection{Target Selection}
Due to telemetry restrictions, less than 200,000 targets had data collected.
In addition, the process of target selection was not random, but was specifically chosen to optimize the detection of Earth-like planets about Sun-like stars.
This biased the target list towards FGK-type main-sequence stars.
Magnitudes and colors obtained from surveys such as SDSS and 2MASS were used to perform the selection.
To simulate the complex target selection process we draw systems from our synthetic sample (single and binary systems) that are as similar as possible, in terms of magnitude and color, to the \textit{Kepler Stellar Properties Catalog} \citep[\kspc, ][]{Mathur.etal.2017}.

The \kspc{} contains, in addition to the broad band visible \kepler{} magnitude, $K_p$, values for the 2MASS infrared $J$, $H$, and $K_s$ bands.
\galaxia{} provides absolute magnitudes for the SDSS $g$ and $r$ bands and the 2MASS $J$, $H$, and $K_s$ infrared bands along with the extinction coefficient $E(B-V)$ for each star.
We compute the apparent magnitudes for each of these bands, while also determining the combined magnitude for each binary treating all binaries as unresolved by \kepler.
To determine the \kepler\ magnitudes we use the Sloan $g$- and $r$-bands to determine $K_p$ via the relationships provided by \citet{Brown.etal.2011}:
\begin{equation}
    K_p = \begin{cases}
        0.2 g + 0.8 r & \text{for } (g - r) \leq 0.8 \\
        0.1 g + 0.9 r & \text{for } (g - r) > 0.8
    \end{cases}
\end{equation}
We determine the synthetic target list using the following merit function:
\begin{equation}
    S_{ts} = \sqrt{
        (K_p^\text{*} - K_p)^2 +
        (J^\text{*} - K_s^\text{*} - J + K)^2 +
        (H^\text{*} - H)^2
    }
\end{equation}
where we denote the parameters from the synthetic population with an asterisk while the parameters without an asterisk are the parameters from the \kspc.
Systems are drawn from the synthetic population, without replacement, for each target in the \kspc.

\begin{figure}
    \includegraphics{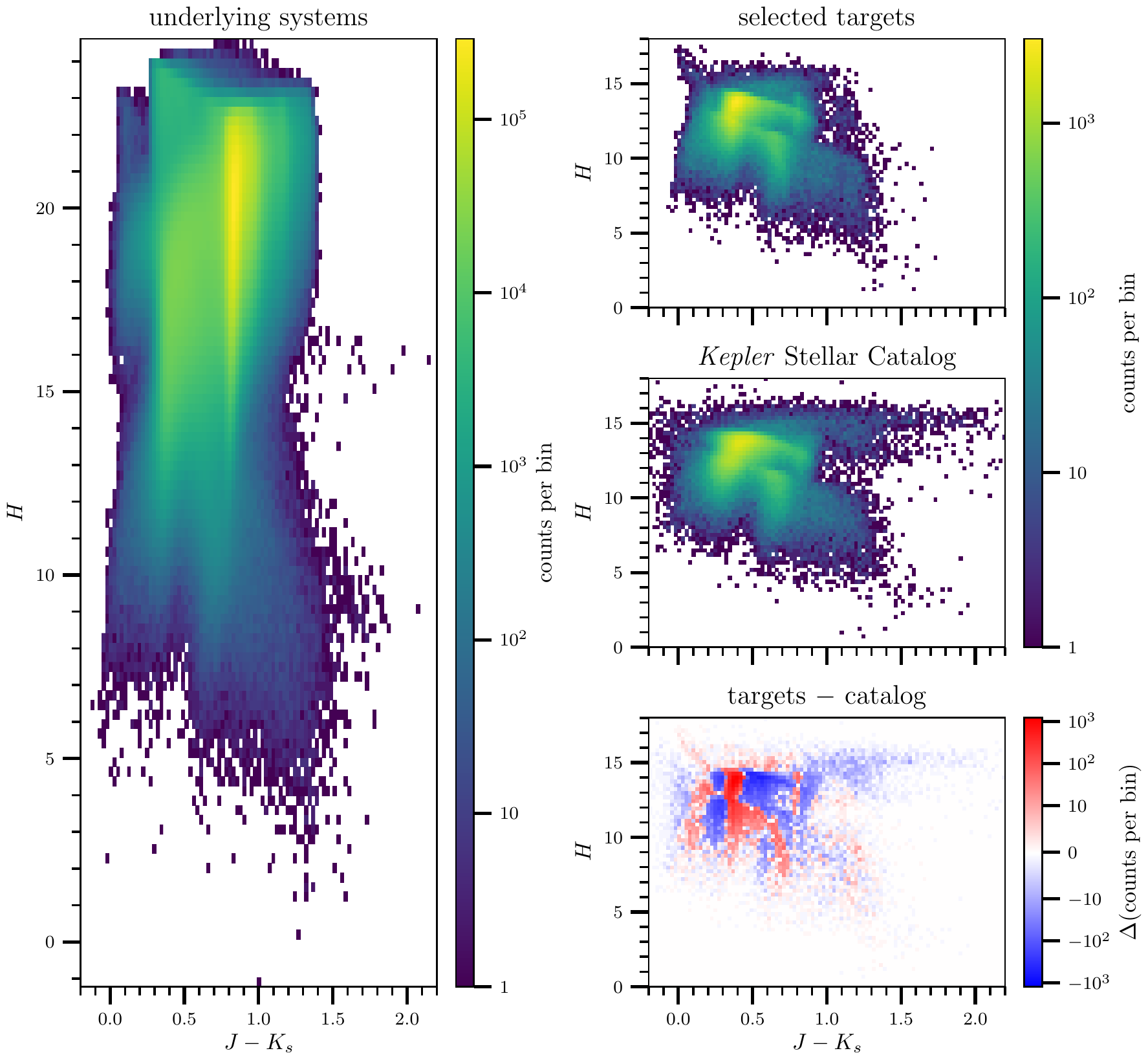}
    \caption{
        The \textbf{left} panel shows the $J-K_s$ versus $H$ distribution of the \rev{synthetic} systems, composed of singles and binaries.
        The \textbf{top right} panel is distribution of the synthetic targets obtained after simulation the selection process.
        The \textbf{center right} panel is the distribution of the \kepler\ Stellar Catalog.
        The \textbf{bottom right} panel shows the difference between the synthetic targets and the actual target list.
        The clusters are the result of the simultaneous constraint on \kepler\ magnitude (see \autoref{fig:kepmag_selection}).
    }
    \label{fig:jhks_selection}
\end{figure}

\begin{figure}
    \includegraphics{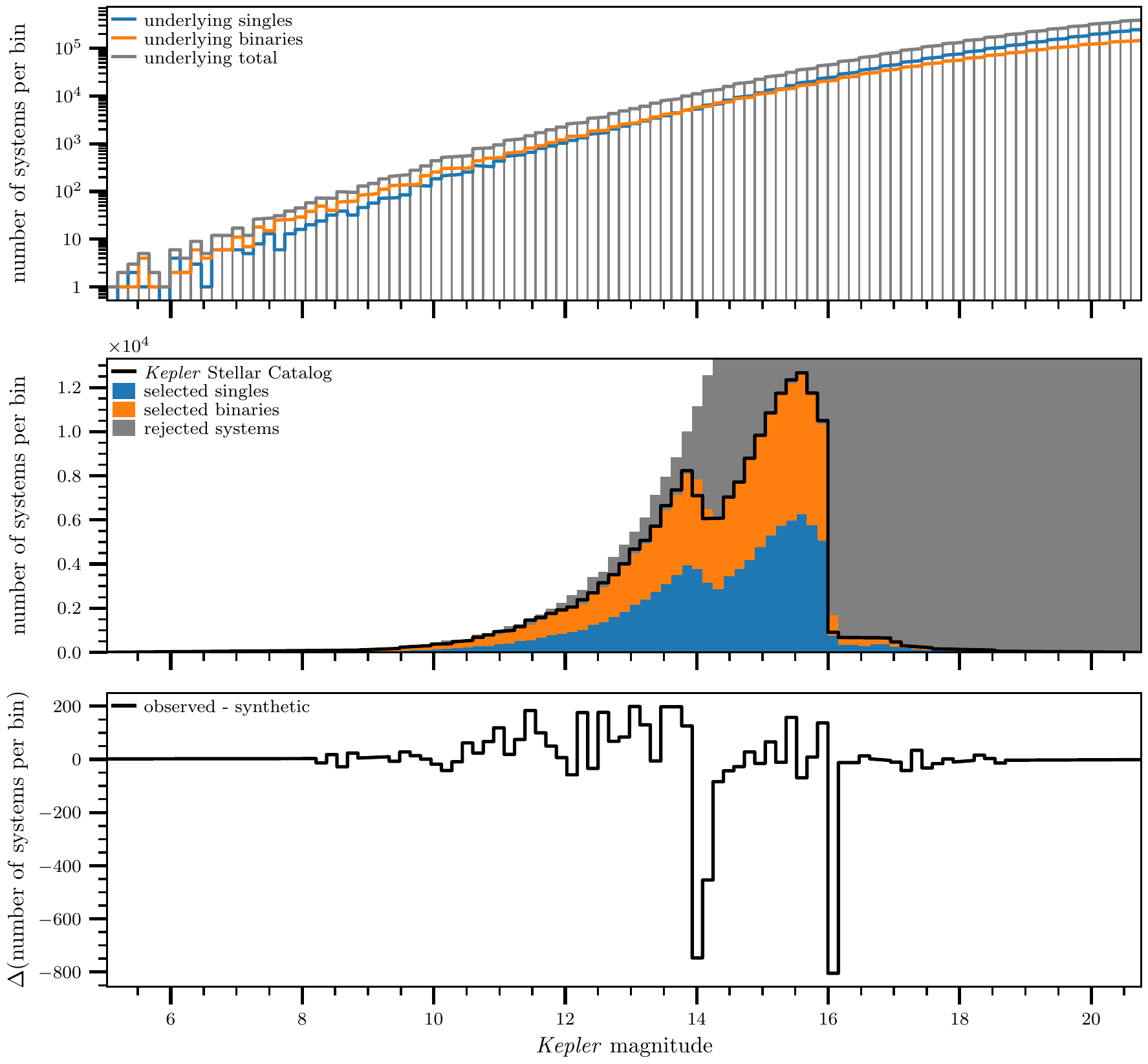}
    \caption{
        In the \textbf{top} panel we show the distribution single systems (in \colorbox{tabblue}{\vphantom{X}}), binaries (in \colorbox{taborange}{\vphantom{X}}), and the resulting total generated as a function of \kepler{} magnitude.
        \rev{The single and binary curves} cross around 14\ssth\ magnitude when more of the underlying systems are comprised of single stars.
        This turnover is due to the fact that the brighter objects tend to be more massive stars and more massive stars have a higher multiplicity fraction than lower mass stars.
        The \textbf{middle} panel shows the fraction of selected systems that are single systems (in \colorbox{tabblue}{\vphantom{X}}) or binaries (in \colorbox{taborange}{\vphantom{X}}) as a function of \kepler\ magnitude.
        In addition, the actual \kepler\ Stellar Catalog (in \textbf{black}) is shown as a reference.
        The \textbf{bottom} panel shows the difference between the \kepler\ Stellar Catalog and the synthetic target sample.
        There are two distinct features at 14\ssth\ and 16\ssth\ magnitude where the synthetic sample slightly over-allocates systems.
        These combined deviations account for approximately 1\% of systems.
        These arise because of the simultaneous constraint on $H$ and $J-K_s$ (see \autoref{fig:jhks_selection}).
    }
    \label{fig:kepmag_selection}
\end{figure}

\rev{The $J-K_s$ versus $H$ distribution} shows the presence of a large population of giants.
In the left panel of \autoref{fig:jhks_selection}, the strip of systems with $J-K_s\approx0.8$ are red giants.
The right middle panel of \autoref{fig:jhks_selection} shows that the catalog contains a small clump of faint red giants, $J-K_s > 0.8$ and $H > 13$, that is not reproduced by our simulated selection process.
Using the conversions provided by \citet{Caldwell1993} and the approximation by \citet{Ballesteros2012}, we find that a star with $J-K_s \approx 1.5$ would have an effective temperature of \SI{3200}{K}.
The relationships by \citet{Caldwell1993} do not hold for values of $J-K_s > 1.5$ but it is safe to say that these would correspond to non-stellar sources.
The relationships used to derive the \kepler{} magnitude have systematic errors as high as 0.6~magnitude toward fainter \kepler{} magnitudes for cooler stars and could account for the clustering features seen in the difference between our synthetic target list and the observed target list.
Looking at the 2 main selection breaks that occur at 14\ssth{} and 16\ssth{} magnitude in \autoref{fig:kepmag_selection}, we can see that our process underselects brighter targets and overselects fainter systems.
For a given $H$ magnitude the corresponding $K_p$ will be overestimated for cooler systems, which explains the discrepancies in \autoref{fig:kepmag_selection}.

\begin{figure}
    \includegraphics{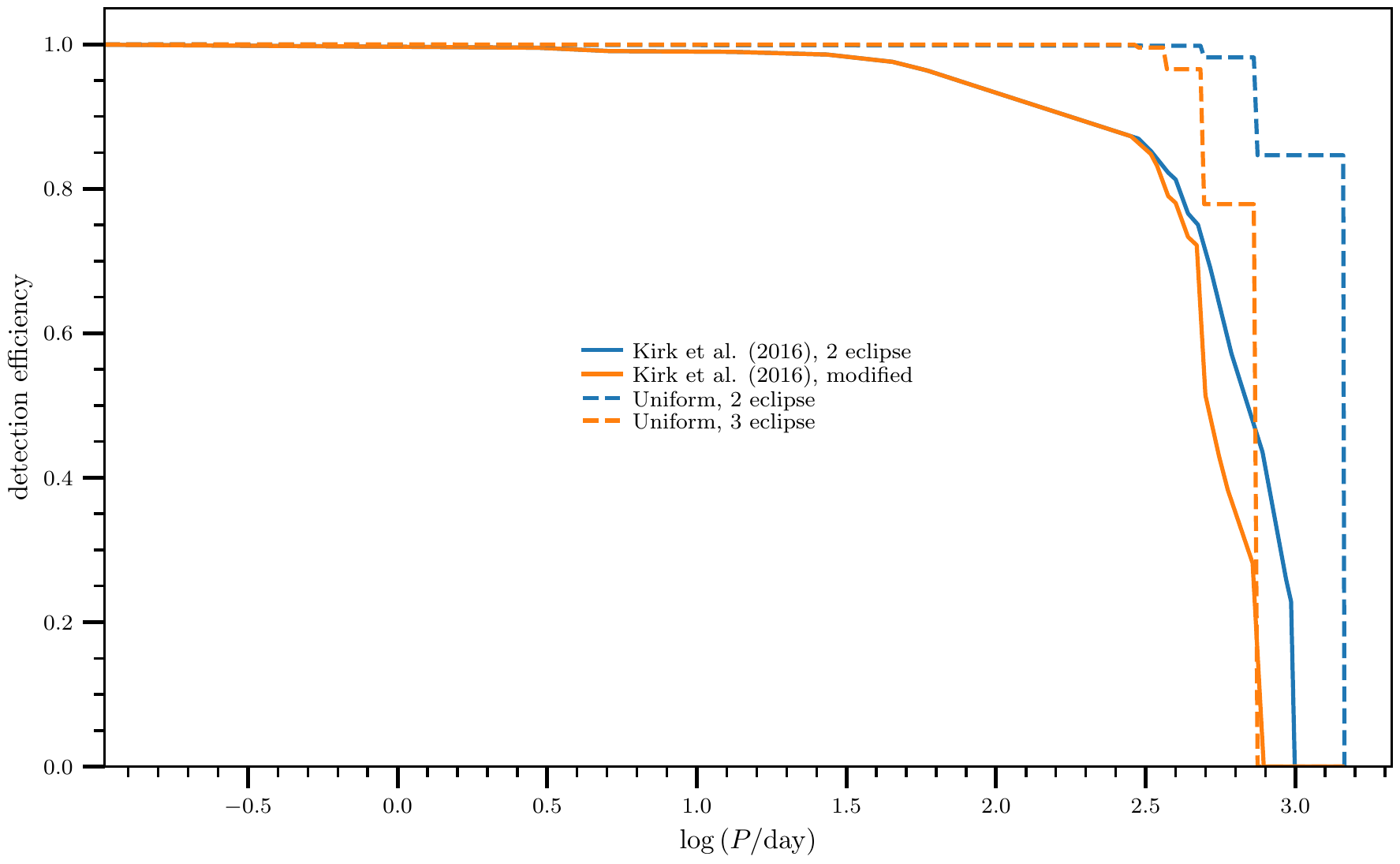}
    \caption{
        The \textbf{dashed blue} and \textbf{dashed orange} lines show the \rev{detection efficiency} for 2 and 3 eclipses, respectively, if a uniform duty cycle of 92\% were assumed.
        The \textbf{solid blue} line is the empirical 2 eclipse \rev{detection efficiency model} reported by \citet{Kirk.etal.2016}.
        The \textbf{solid orange} line is the 2-eclipse \rev{detection efficiency relationship} modified for 3 eclipses.
        The cutoff at $\logp=2.86$ corresponds to the upper limit of \SI{730}{d}.
    }
    \label{fig:dutycycle}
\end{figure}

\rev{\subsection{Detection Efficiency}}
The instrument also suffered from gaps in the data.
Some of these gaps were scheduled, such as rolling of the spacecraft to reorient the solar panels toward the Sun and to download data, while others, like the failure of two CCDs, were not.
An empirical 2-eclipse \rev{detection efficiency model} is provided by \citet{Kirk.etal.2016}, see \autoref{fig:dutycycle}.
Their empirical \rev{detection efficiency} model was obtained by scanning all of the \kepler{} lightcurves over a range of observable test periods.
We modify their empirical relationship by first considering the scenario of a \rev{periodic} duty cycle.

The probability of observing exactly $k$ eclipses out of $N$ eclipsing events is given by the binomial distribution
\begin{equation*}
    p_k = \binom{N}{k}f_\textnormal{dc}^k(1-f_\textnormal{dc})^{N-k}
\end{equation*}
where $f_\textnormal{dc}$ is the duty cycle.
The number of eclipses that can occur for a given period is
\begin{equation*}
    N=1+\lfloor(\SI{1460}{d})/P\rfloor
\end{equation*}
where \SI{1460}{d} is the length of the original \kepler\ mission.
The \kebc\ only includes systems that have had three eclipses observed.
The probability of at least 3 observed eclipses is the same as 
\begin{equation*}
    p_{k\ge3} = 1 - p_0 - p_1 - p_2
\end{equation*}
where $p_0$, $p_1$, and $p_2$ are the probabilities of observing 0 eclipses, only 1 eclipse, and exactly 2 eclipses, respectively.
Our modified 3 eclipse \rev{detection efficiency} function is, using $f_\textnormal{dc}=0.92$, is 
\begin{equation*}
    p_{k\ge3} = f_\textnormal{emp} - 0.5N(N-1)(0.92)^2(0.08)^{(N-2)}
\end{equation*}
where we have replaced the uniform probability of observing at least 2 eclipses with the empirical relationship from \citet{Kirk.etal.2016}, $f_\textnormal{emp}$.


\section{Calibration} \label{sec:calibration}
Once a synthetic eclipsing binary sample has been computed, we compare its \logp\ distribution to the observed \logp\ distribution from the \kebc{}.
The \kebc\ only has published values for periods but future work will expand this process to the mass ratio and eccentricity distributions.
We create a discrete \logp{} pdf, $\mathcal{P}_i$, where the $i$\ssth{} value is the probability density for the $i$\ssth{} bin.
\rev{%
A relaxation method is used to update the pdf by comparing the histograms of the observed and synthetic period distributions.
}%
The relative difference for the $i\ssth{}$ bin, in counts, is computed as:
\begin{equation}
    \Delta_i = \frac{C_{i,K} - C_{i,S}}{C_{i,K}}
    \label{eqn:delta}
\end{equation}
where $C_{i,K}$ and $C_{i,S}$ are the number of eclipsing binaries in the $i\ssth{}$ bin for the \kebc\ and the synthetic eclipsing binary catalog, respectively.
We compute the updated value for each bin, $\mathcal{P}_i^\text{*}$, via
\begin{equation}
    \mathcal{P}_i^\text{*} = \mathcal{P}_i (1 + r\Delta_i^\mathrm{corr})
    \label{eqn:update}
\end{equation}
where $r$ is the \rev{relaxation rate} and the corrected difference term is
\begin{equation*}
    \Delta_i^\mathrm{corr} = \sgn(\Delta_i)\min(1, \lvert \Delta_i \rvert).
\end{equation*}
The \rev{relaxation rate} takes on values between 0 and 1 and for this work we adopt $r=0.3$ as the optimal value.
Setting the \rev{relaxation rate} too low will stall convergence while setting it too high will cause the model to wildly oscillate.
The corrected difference term, $\Delta_i^\mathrm{corr}$, guarantees that the right-hand side of \autoref{eqn:update} is greater than zero.
Without $\Delta_i^\mathrm{corr}$, there is nothing keeping the bins from assuming negative values.

In addition, we append another bin, $\mathcal{P}_\mathrm{out}^\text{*}$, to the end of the distribution that is outside of the observable range of periods.
This bin is computed via
\begin{equation}
    \mathcal{P}_\mathrm{out}^\text{*} = \frac{1}{w_\mathrm{out}}\left(1 - \sum_i\mathcal{P}_i^\text{*}w_i\right)
\end{equation}
where $w_i$ and $w_\mathrm{out}$ are the associated bin widths.
The $\mathcal{P}_\mathrm{out}^\text{*}$ bin functions as a normalizing constant by providing a mechanism to fit the absolute number of systems observed.
If we only generated periods within the \kepler\ observable period window we would greatly overestimate the number of observed eclipsing systems.

The initial shape of the period distribution only impacts the rate of convergence.
We use a decaying exponential to accelerate the convergence of the upper end of the period distribution; we are able to achieve convergence after 300 iterations.
In contrast, if we start with a flat distribution over 1000 iterations are required.
As long as the initial value of each bin is non-zero the converged solution is the same.
Care must be taken when initializing the bins as setting bins to zero (or very close to zero) will prevent them from being adjusted as the level of adjustment is based on the magnitude of the bin's previous value per \autoref{eqn:update}.


\section{Results}

\begin{figure}
    \centering
    \includegraphics{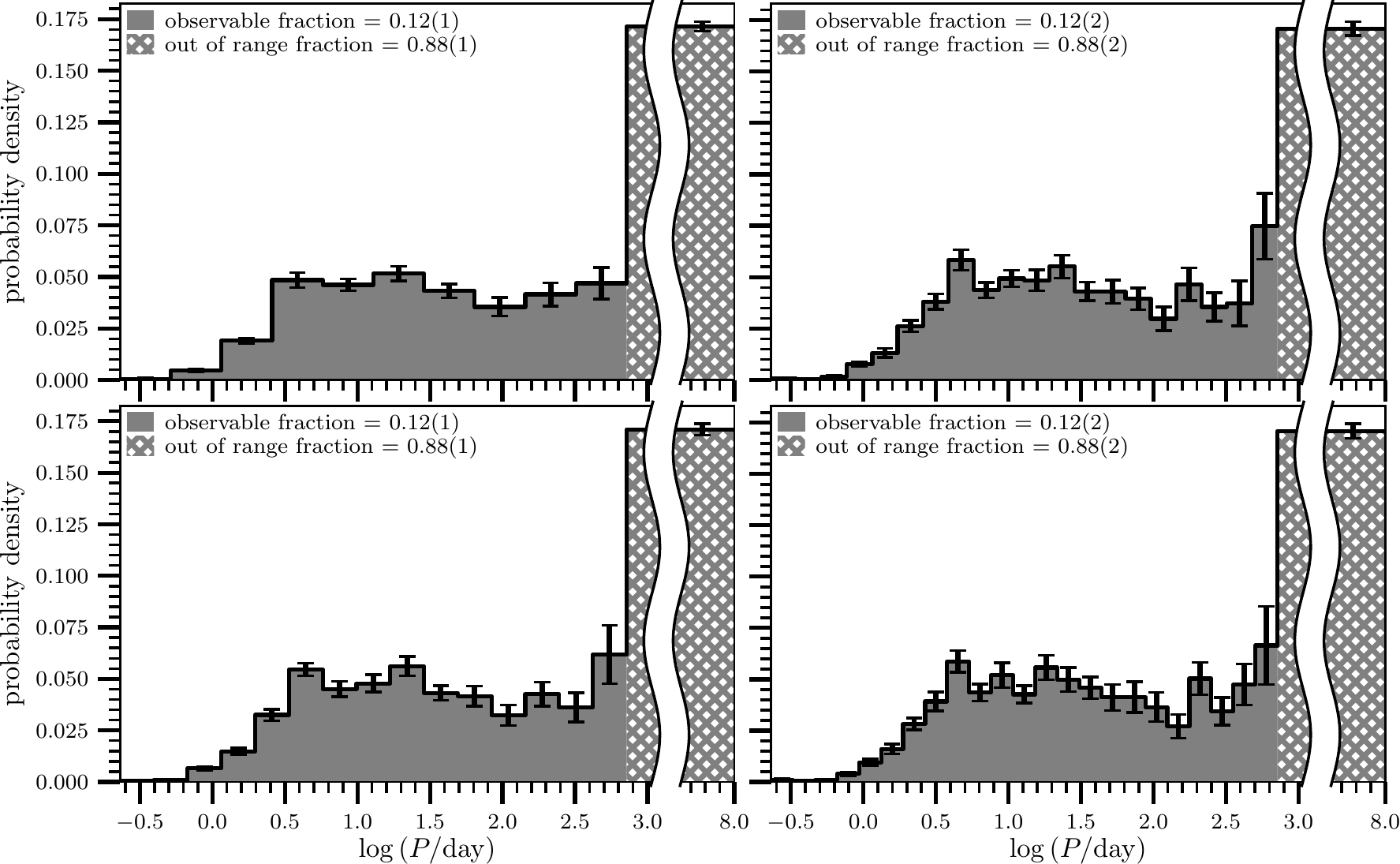}
    \caption{
        The resulting \logp\ probability density function for \rev{the \kepler\ binary population}.
        We present the results of four different binning strategies were we have only varied the number of bins that cover the \kepler\ observable period range: \textbf{top left} uses 10~bins, \textbf{bottom left} uses 15~bins, \textbf{top right} uses 20 bins, and \textbf{bottom right} uses 23 bins.
        The overflow bin is visually truncated, as represented by the vertical breaks, for convenience.
    }
    \label{fig:logp_models}
\end{figure}

The number of bins used in our discrete model can induce spurious structure into the resulting population.
Too few bins leads to poor resolution while too many bins can produce overfitting.
The rule given by \citet{Freedman1981} suggests using 23 bins across the observable period range.
We used four binning strategies and present the resulting probability density functions in \autoref{fig:logp_models}.
The integrated \logp\ distribution over the observable range is 0.12 for each choice of binning.
In addition, the distributions are all in agreement within their 1~sigma uncertainties.
The 20 and 23 bin models result in excessively noisy solutions that appear to overfit the data.
We will present the results of the model using the 15 bin version of the \logp{} distribution.

\subsection{Calibration}
\begin{figure}
    \includegraphics{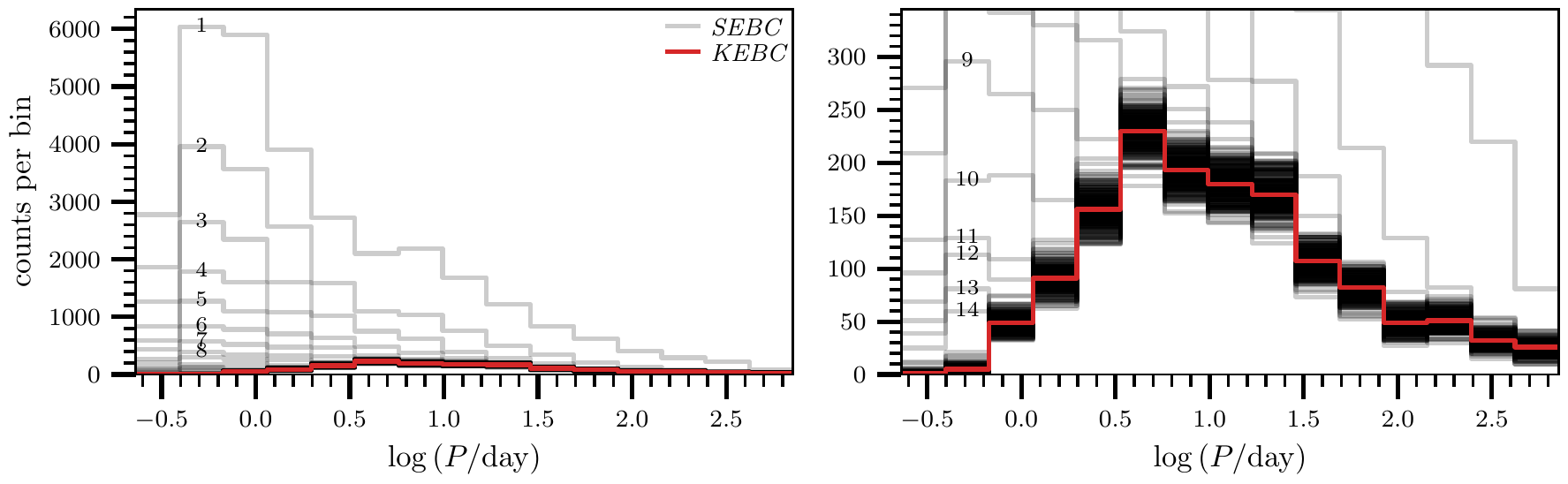}
    \caption{
        The observed synthetic eclipsing binary catalog (\sebc), in \textbf{semi-transparent black}, after each iteration plotted against the \kebc{}, in \textbf{red}.
        The first few iterations have been labeled in order.
        The \textbf{right} plot provides a zoomed version of the \textbf{left}.
        The darker regions are areas where multiple iterations are overlapping.
    }
    \label{fig:calibration_15}
\end{figure}

The process of fitting the \logp\ distribution begins by vastly overestimating the total number of systems.
\autoref{fig:calibration_15} shows the synthetic eclipsing binary catalog (\sebc) after each run.
Each bin is updated after every iteration and after roughly 15~runs the total number of synthetically observed systems matches the total number from the \kebc.
We run 250~iterations to ensure that the model has burned in about the solution.
To get our resulting model, we perform an additional 50~iterations from which we compute the mean bin values and associated uncertainties.

\subsection{Binary Population Parameters}
\begin{figure}
    \includegraphics{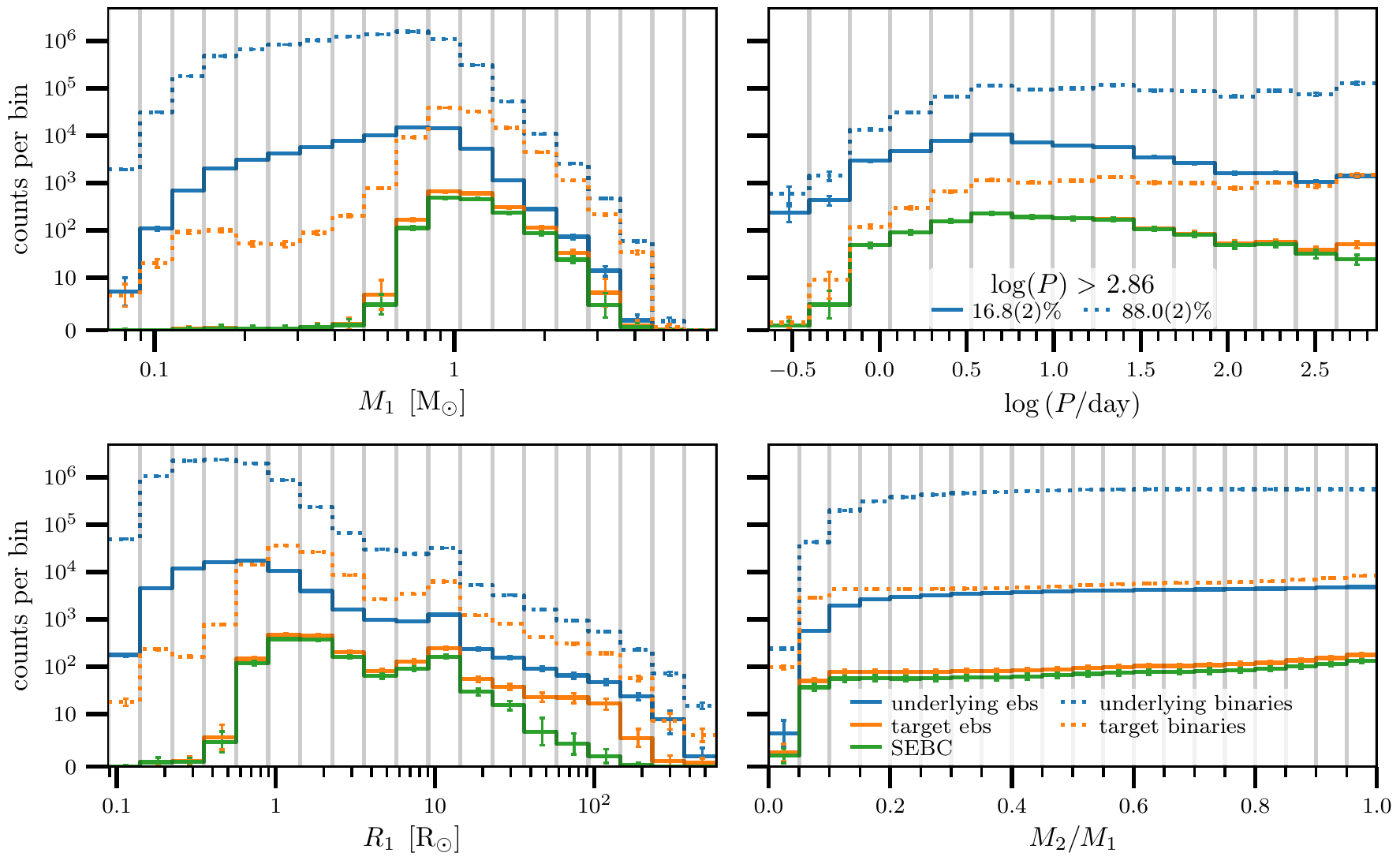}
    \caption{
        The resulting model distributions, generated using the 15~bin version of the \logp\ distribution, are plotted with primary mass in the \textbf{top left}, primary radius in the \textbf{bottom left}, \logp\ in the \textbf{top right}, and the mass ratio in the \textbf{bottom right}.
        For each parameter distribution we show the \rev{simulated} underlying binary population (\textbf{\textcolor{tabblue}{dotted blue}}), the target-selected sample of binaries (\textbf{\textcolor{taborange}{dotted orange}}), \rev{the eclipsing} binary population (\textbf{\textcolor{tabblue}{solid blue}}), the target-selected eclipsing binary sample (\textbf{\textcolor{taborange}{solid orange}}), and the synthetic eclipsing binary catalog (SEBC, \textbf{\textcolor{tabgreen}{solid green}}).
        \rev{The SEBC will be directly compared against the \kebc{}.}
        For the \logp\ distribution we estimate that 88.0(2)\% of the underlying binaries and 16.8(2)\% of the underlying eclipsing binaries have periods higher than $\logp{}>2.86$.
    }
    \label{fig:mass_radi_logp_m2m1}
\end{figure}

The resulting binary parameter distributions are shown in \autoref{fig:mass_radi_logp_m2m1}.
\rev{The mass distribution} shows a decline at low masses.
This drop-off is a result of using a magnitude-limited sample.
The mass distribution of the target-selected sample peaks around a solar mass.
This is in line with the target selection process and shows that our process of simulated target selection captures the distribution of targets appropriately.
The selected eclipsing population is dominated by binaries with $M_1 > \SI{0.5}{\msun}$.
This is due to a combination of effects.
The first is the declining fractional eclipse probability for smaller, and typically less massive stars, which can be seen if one compares the shape of \rev{the eclipsing binaries to the binaries}.
Eclipse probability is proportional to the sum of fractional radii so smaller stars will present eclipses less frequently than larger stars.
When combined with the target selection bias, eclipsing binaries below \SI{0.5}{\msun} and \SI{0.4}{\rsun} are effectively suppressed.

The actual target list was not free of giants, but included roughly 5000 stars with $\log{g} < 3.5$.
The synthetic target-selected radius distribution is in agreement having less than 2000 binaries with $R_1 > \SI{10}{\rsun}$.
There appears to be a discrepancy between the target-selected eclipsing binary sample and the observed eclipsing binaries which widens for larger stars.
This follows directly from our minimum period constraint in \autoref{eqn:r1r2}: systems with larger stars are more likely to have values of $\logp{} > 2.86$ and are unable to have 3~eclipses observed.

\rev{The \logp\ distribution} appears to be relatively flat above 0.5 (roughly \SI{3.2}{d}).
We estimate that the period range covered by \kepler\ contains 12.0(2)\% and 83.2(2)\% of all binaries and eclipsing binaries in the field, respectively.
\rev{The fraction of eclipsing binaries to binaries} declines at longer periods because of the larger separation of the system which reduces eclipse efficiency.
The discrepancy between the selected and observed eclipsing binary samples at longer \logp\ is due to the \rev{lower detection efficiency}.
The obtained mass ratio distribution, $M_2/M_1$, reflects the fixed target mass ratio distribution $q$.

\begin{figure}
    \includegraphics{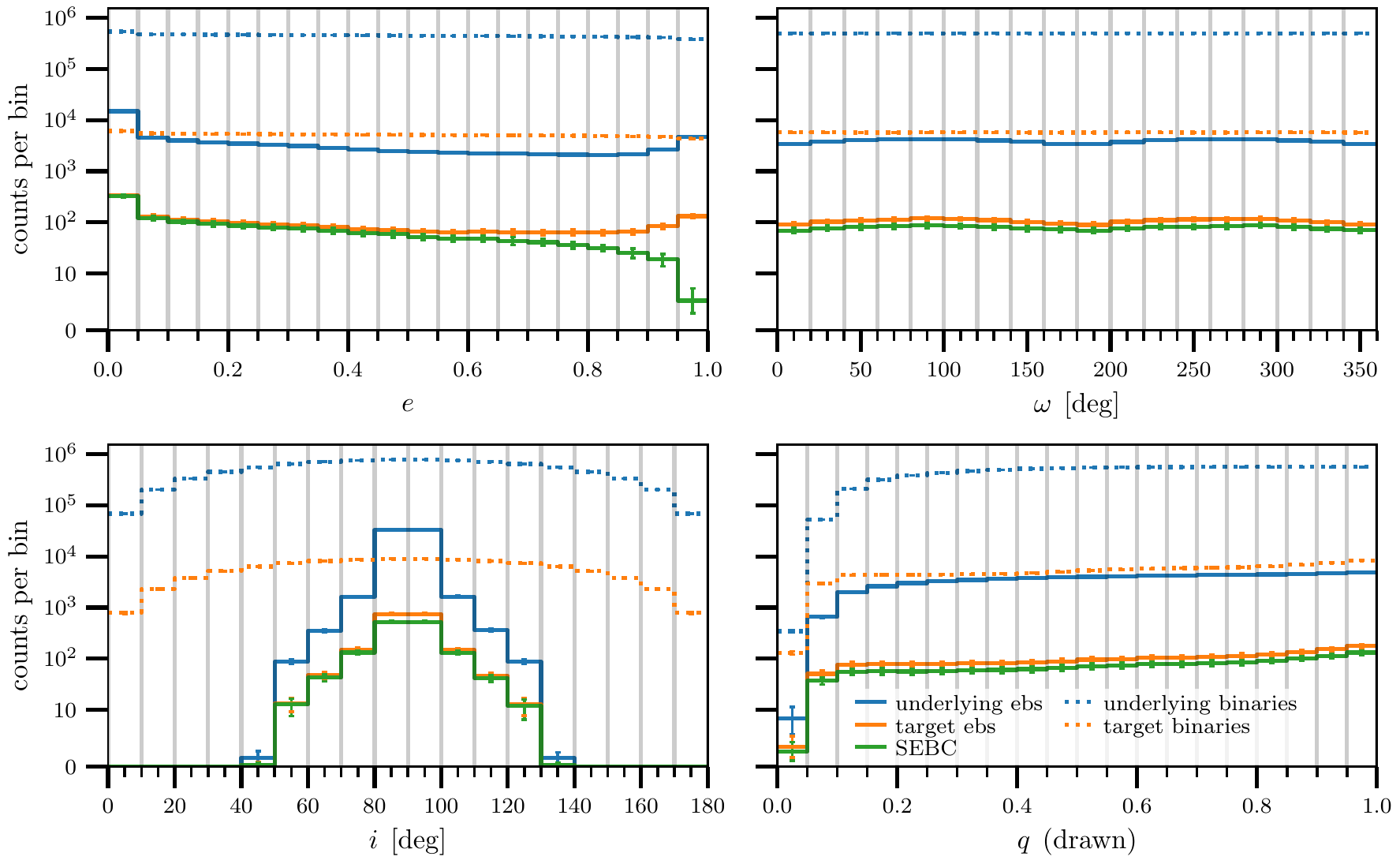}
    \caption{
        The fixed model distributions, generated using the 15~bin version of the \logp\ distribution, are plotted with eccentricity in the \textbf{top left}, inclination in the \textbf{bottom left}, argument of periastron in the \textbf{top right}, and the target mass ratio in the \textbf{bottom right}.
        For each parameter distribution we show the \rev{simulated} underlying binary population (\textbf{\textcolor{tabblue}{dotted blue}}), the target-selected sample of binaries (\textbf{\textcolor{taborange}{dotted orange}}), \rev{the eclipsing} binary population (\textbf{\textcolor{tabblue}{solid blue}}), the target-selected eclipsing binary sample (\textbf{\textcolor{taborange}{solid orange}}), and the synthetic eclipsing binary catalog (\textbf{\textcolor{tabgreen}{solid green}}).
    }
    \label{fig:eccn_incl_argp_mrat}
\end{figure}

\autoref{fig:eccn_incl_argp_mrat} shows the input distributions that were held fixed throughout the synthesis.
\rev{The eccentricity distributions of the simulated underlying binary population and target-selected binary sample show a slight abundance in the first bin due to circularization.}
\rev{These distributions} show that eclipse detection is boosted for circularized systems and high eccentricity systems.
Circularized systems tend to have shorter periods than eccentric systems and thus have a larger eclipse detection efficiency.
High eccentricity systems will have a boost in detection efficiency because of the closer approach of the systems near periastron.
The discrepancy in the eccentricity distributions for selected eclipsing binaries and the observed eclipsing binaries is related to the constraint from \autoref{eqn:r1r2}.
Higher eccentricities require longer periods making these systems more likely to have values of \logp\ outside of the observable window.
The decline of \rev{the number} of systems for smaller mass ratios is due to the minimum mass ratio constraint provided in \autoref{eqn:q0}.
In addition, smaller and larger mass ratios are over-represented in the target list as compared to more moderate mass ratios.
This is due to the shape of the mass ratio distribution with respect to primary mass, presented \autoref{fig:mass-mrat}.
Smaller mass ratios are going to be present only in systems with more massive primaries making these more likely to make it into the target list.
Higher mass ratio systems will tend to be brighter and will be slightly favored for a given primary mass.

The $i$ and $\omega$ parameters pertain to the viewing angle and show no unexpected features for the binary population.
The $i$ distribution shows that eclipses only occur between \SI{45}{\degree} and \SI{135}{\degree} which is precisely what one would expect for spherical stars.
The $\omega$ distribution shows that eclipses are more likely when periastron is aligned with the plane of the sky, which occurs at \SI{90}{\degree} and \SI{270}{\degree} as opposed to when periastron is perpendicular to the plane of the sky, which occurs at \SI{0}{\degree} and \SI{180}{\degree}.

\subsection{\rev{Validation}}

\begin{figure}
    \centering
    \includegraphics{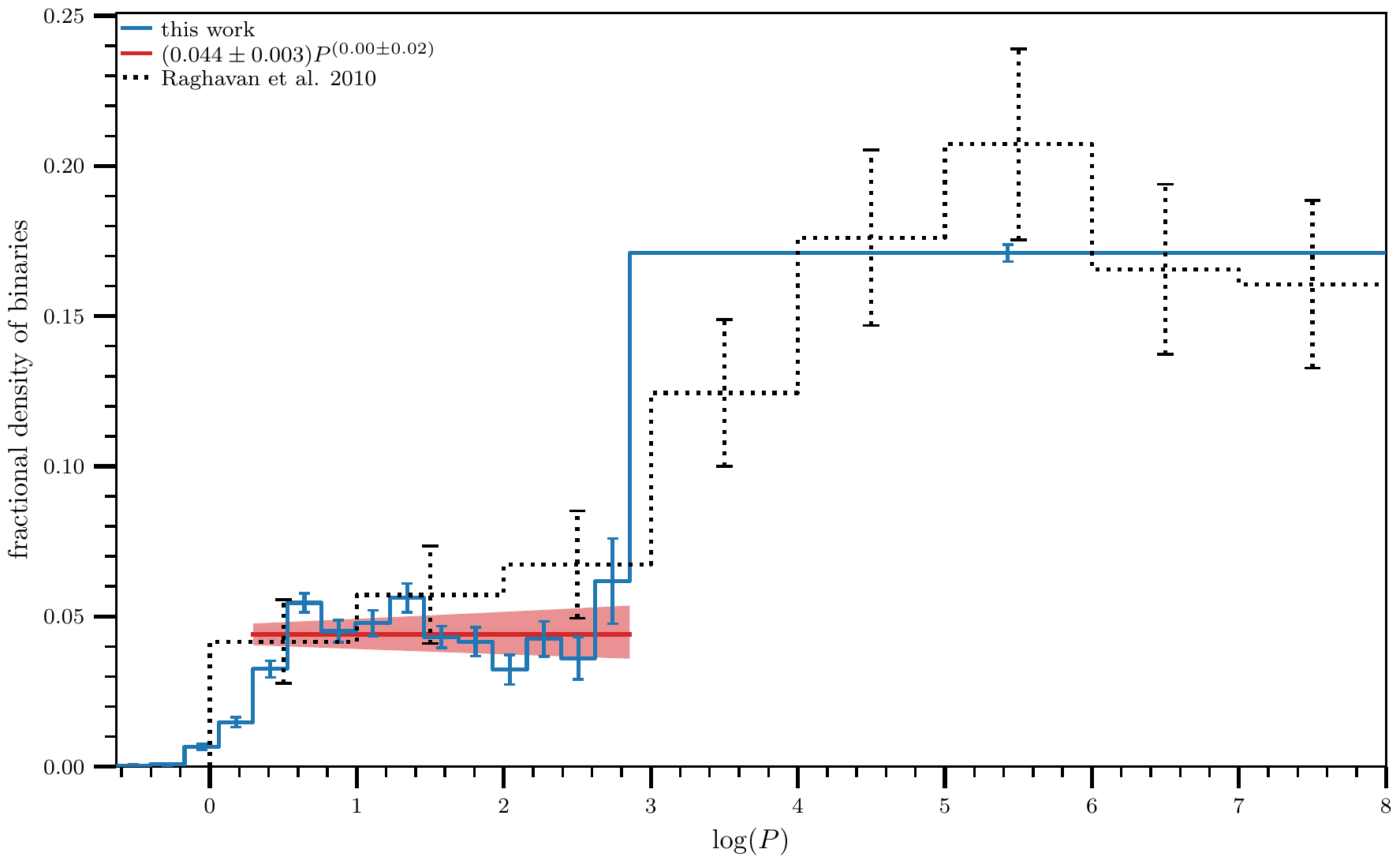}
    \caption{\rev{%
        Comparison of the \logp{} distribution from our model (\textbfcol{tabblue}{blue}) and the volume-limited sample of solar-like stars (\textbf{dotted}) from \citet{Raghavan.etal.2010} as reported by \citet{Moe.DiStefano.2017}.
        In addition, a power law relationship (\textbfcol{tabred}{red}) has been fit to data from \SI{2}{d} to \SI{730}{d}. The shaded red region corresponds to 1$\sigma$ uncertainty determined by the Markov Chain Monte Carlo (MCMC) sampler.
    }}
    \label{fig:logp_result}
\end{figure}

\rev{%
In order to validate our resulting \logp{} distribution, we compare it to the volume-limited sample of \citet{Raghavan.etal.2010}, as reported in \citet{Moe.DiStefano.2017}.
While the \citet{Raghavan.etal.2010} sample is composed of solar-like stars found within \SI{25}{pc} about the Sun, it still provides a useful point of comparison for our results.
\autoref{fig:logp_result} shows that the \logp{} distribution of our model is consistent with \citeauthor{Raghavan.etal.2010} within uncertainty.
}%
\rev{%
A power law of the form $f(P)=kP^\alpha$ was fit to the derived distribution over the range of \SI{2}{d} to \SI{730}{d}, with $k=0.044\pm0.003)$ and $\alpha=0.00\pm0.02$.
This range was chosen to avoid circularization effects (cf.~\autoref{eqn:emax}) at shorter periods.
The resulting fit shows a flat curve consistent with \citet{Raghavan.etal.2010} within 1$\sigma$ uncertainty.
}%

\subsection{Sensitivity}
\begin{figure}
    \centering
    \includegraphics{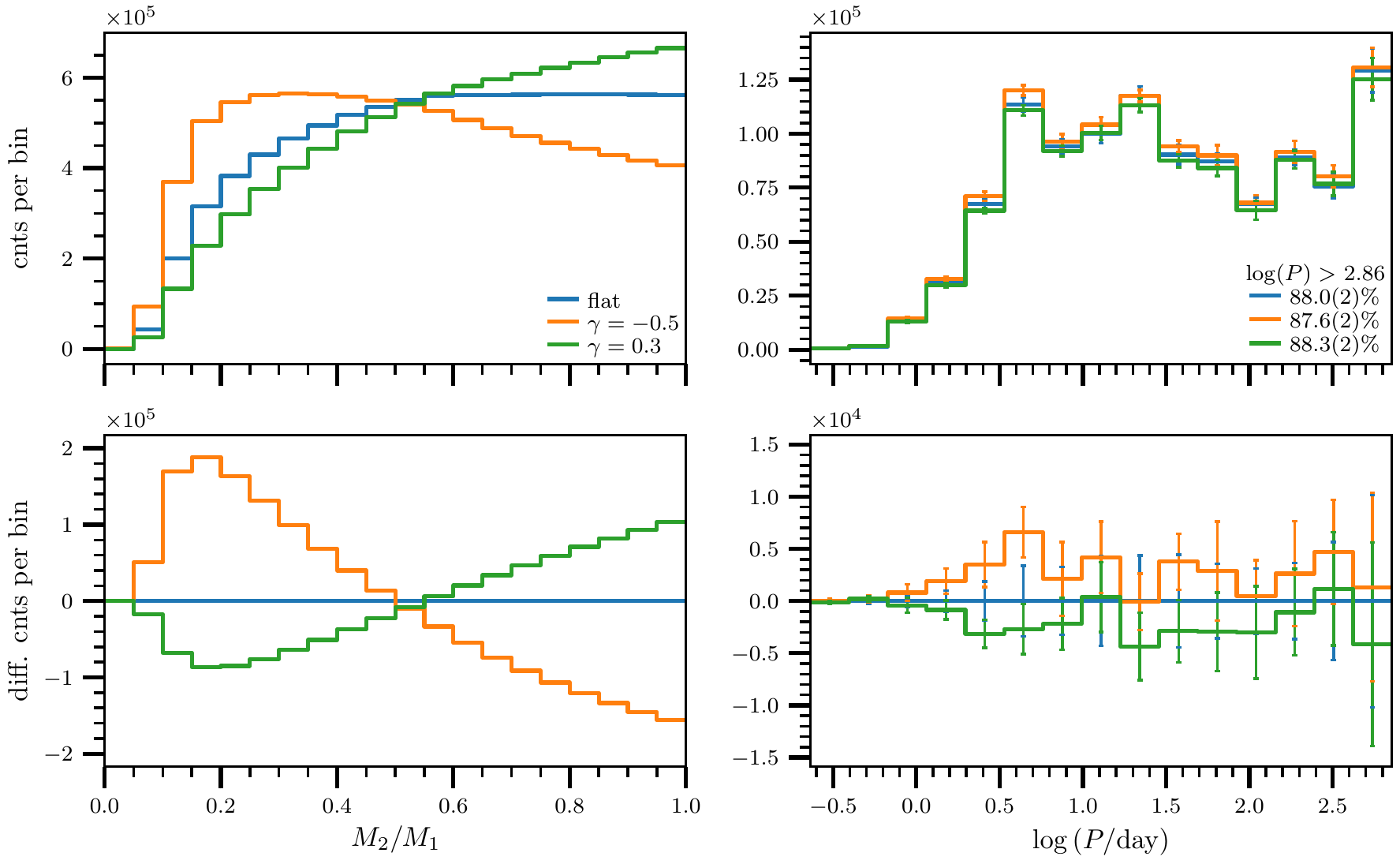}
    \caption{
        These plots show how sensitive the \rev{resulting} period distribution is to the \rev{input} mass ratio distribution.
        In the \textbf{top left} we plot three mass ratio distributions: flat (\textbfcol{tabblue}{blue}), $f(q)\propto q^{-0.5}$ (\textbfcol{taborange}{orange}), and $f(q)\propto q^{0.3}$ (\textbfcol{tabgreen}{green}).
        The \textbf{top right} plot shows the corresponding \logp\ distributions and its overflow bin percentage.
        The \textbf{bottom left} plot shows the difference between each mass ratio distribution and the flat distribution.
        The \textbf{bottom right} plot shows the difference plot between each \logp\ distribution and the \logp\ distribution obtained from a flat mass ratio distribution.
    }
    \label{fig:m2m1_logp_comp}
\end{figure}

\begin{figure}
    \centering
    \includegraphics{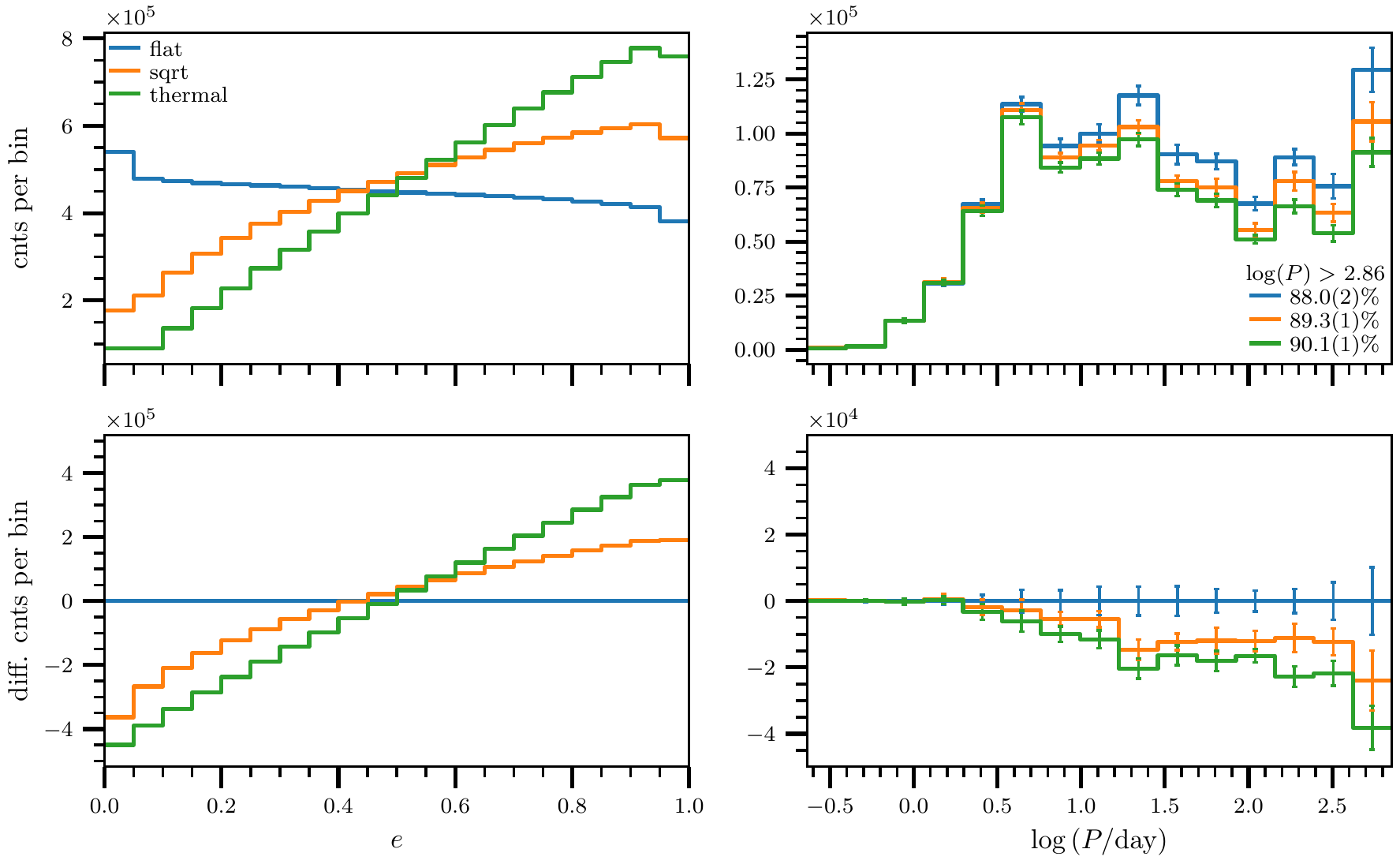}
    \caption{
        These plots show how sensitive the \rev{resulting} period distribution is to the \rev{input} eccentricity distribution.
        In the \textbf{top left} we plot three eccentricity distributions: flat (\textbfcol{tabblue}{blue}), $f(e)\propto \sqrt{e}$ (\textbfcol{taborange}{orange}), and thermal (\textbfcol{tabgreen}{green}).
        The \textbf{top right} plot shows the corresponding \logp\ distributions and its overflow bin percentage.
        The \textbf{bottom left} plot shows the difference between each eccentricity distribution and the flat distribution.
        The \textbf{bottom right} plot shows the difference plot between each \logp\ distribution and the \logp\ distribution obtained from a flat eccentricity distribution.
    }
    \label{fig:eccn_logp_comp}
\end{figure}

\begin{figure}[htb!]
    \includegraphics{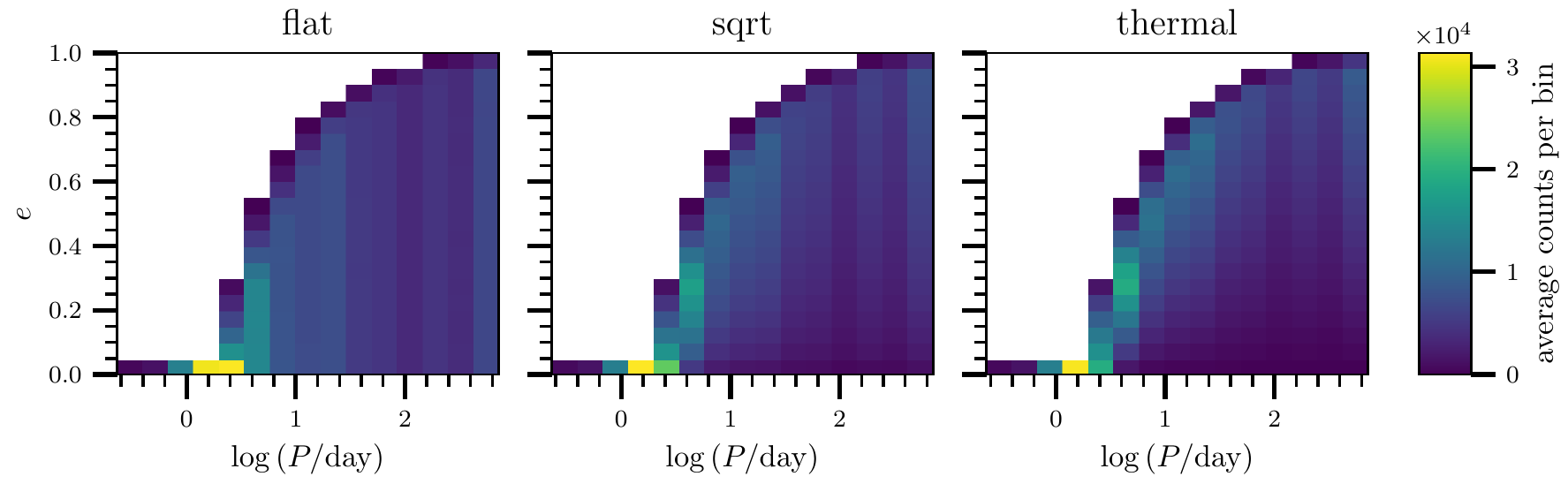}
    \caption{
        Histograms of the \rev{resulting} \logp{}-\eccn{} relationship from different input eccentricity distributions and the resulting \logp\ distribution.
        The histogram for a flat eccentricity distribution (\textbf{left}) shows very little variation in the vertical direction.
        The sqrt distribution (\textbf{middle}) shows an increase in systems for a given period with eccentricity.
        The thermal distribution (\textbf{right}) shows the same trend, except more pronounced.
        The upper left portion of each plot is forbidden by \autoref{eqn:emax}.
    }
    \label{fig:logpeccn}
\end{figure}

We investigated how our choice of input distributions, namely mass ratio and eccentricity, affected \logp{}.
\rev{\autoref{fig:m2m1_logp_comp} shows three separate input mass ratio distributions and the resulting \logp\ distribution for each.}
We conclude from the difference plot that the resulting \logp{} distribution is only marginally coupled to the mass ratio distribution (up to $2\sigma$ deviations).
The overflow bin percentage also shows the relatively low coupling between mass ratio with less than a percent variation across all three distributions.
\autoref{fig:eccn_logp_comp} shows the three separate \rev{input eccentricity distributions and the resulting \logp\ distribution for each.}
Steeper eccentricity distributions result in an attenuation of the \logp\ distribution in the observable period range by shuffling the systems to even longer periods.
The difference plot shows that these eccentricity distributions do result in distinct period distributions.
\autoref{fig:logpeccn} shows the \rev{simulated} underlying binary distribution of $e$ as a function of \logp{}.
These distributions have identical boundaries because of the restriction on \emax{}.
The non-flat distributions contain very fewer low eccentricity systems above $\logp{} > 0.5$.
Ongoing work (Pr\v{s}a et al.) will provide accurate eccentricities for the majority of the \kebc{}.

\section{Discussion}

In this work we outline the basic framework of our binary population synthesis framework and use it to fit a model of the binary population of the \kepler{} field.
Binary systems are constructed from a synthetic stellar population in accordance with our assumed form of the underlying orbital parameters along with imposed constraints.
We compute the observing geometries of these systems to obtain a sample of eclipsing binaries.
The process of target selection is simulated and we correct for the \kepler\ \rev{detection efficiency} to obtain a synthetic eclipsing binary catalog.
This generated catalog of eclipsing binaries is then compared to the actual observed distribution of eclipsing binaries in the \kebc.
We apply corrections to \rev{the binary} distribution and iterate this process until the resulting simulated eclipsing samples simply fluctuate about the observed \kebc{}.
The result is a model of the underlying binary population.

There are several aspects that could be improved upon.
For this work we have not considered the eclipse signal strength during our simulation.
This would require a \kepler\ noise model to see if an eclipse of a certain duration would be detectable.
To get a measure of the signal produced by an eclipsing binary, a limb darkening model would be required to properly compute the eclipse depth.
In addition, eccentric binaries could cause only one component to be eclipsed.
With only access to the light curve, these single eclipsing systems could cause period confusion as the signal could come from a circular system with nearly identical components.
We will continue to explore ways to more realistically model these non-trivial aspects.

Future work will include the reduction of spectra to obtain additional reference distributions used in the forward-model.
We have follow-up spectroscopic observation for 611~objects in the \kebc{} and anticipate that many, if not most, of these objects will allow for radial velocities of one or both components to be measured.
For systems with measurable radial velocity curves for both components, it is possible to not only estimate the separation and eccentricity but also the mass of both stars.

When combined with parallaxes from \gaia{}, \kepler{} light curves can be converted to flux in absolute units.
With a handle on the geometry of the system as well as the surface brightness of each star, a model of the binary system can be fit to the light curve (as well the radial velocity curve) to obtain estimates for the radius and effective temperate of both stars along with the orbital parameters.
These spectra have the potential to yield masses, mass ratios, and eccentricities for a considerable fraction of the \kebc{}.
With these additional reference distributions we can extend the fitting beyond only the period distribution.

This forward-modeling framework is not limited to \kepler{} but can be applied and adapted to future survey missions.
These upcoming missions, including \lsst{}, will observe large populations of eclipsing binaries across the entire sky enabling the inclusion of other relationships such as galactic latitude.
Not only will \lsst{} yield a larger observed sample of eclipsing binaries, but it will observe a more diverse binary population.

\software{Galaxia (Sharma et al. 2011), K2fov (Mullally et al. 2016)}
\medskip

\bibliography{main}

\end{document}